\documentclass[10pt]{article}
\usepackage{lineno}
\usepackage[english]{babel}
\usepackage{amsmath}
\usepackage{placeins}
\usepackage{amssymb}
\usepackage{latexsym}
\usepackage{amsmath}
\usepackage{amsthm}
\usepackage{enumerate}
\usepackage{epsfig}
\usepackage{graphicx}

\usepackage{SProcs}

\begin{document}


\setcounter{page}{1091}
\setcounter{pagelen}{6}
\Stitle{Improvement of random LHD for high dimensions}

\Sauthor{Andrey Pepelyshev\index{Pepelyshev A.}\footnote{University of Sheffield, 
E-mail: \texttt{a.pepelyshev@sheffield.ac.uk}}}

\Smaketitle
\Saddcontentsline{Pepelyshev A.}

\vspace{5mm}

\begin{abstract}
Designs of experiments for multivariate case are reviewed. 
Fast algorithm of construction of good  Latin hypercube designs is developed.
\end{abstract}

\section{Introduction}

The mathematical theory for designing experiments was started to develop by Sir Ronald A. Fisher
who pioneered the design principles in his studies of analysis of variance
originally in agriculture. The theory of experimental designs have received  considerable development further in
the middle of the twentieth century in works by G.E.P Box, J. Kiefer and many others.
Computer experiments have become available with the appearance of computer engineering.
Mathematical computer models are a replacement for natural (physical, chemical, biological)
experiments which are too time consuming or too costly. Moreover, mathematical models may describe 
phenomena which can not be reproduced, for example, weather modeling.

Experimental designs for deterministic computer models was studied  first by McKay et al. (1979).
The theoretical principles of analysis of deterministic computer models were
determined in Sacks et al. (1989) and the analysis of simulation models 
(deterministic computer codes with stochastic output) in Kleijnen (1987).
During the last decade the Bayesian approach to computer experiments was extensively developed,
see Kennedy, O'Hagan (2001), Conti, O'Hagan (2008) and references within.
The technique used in the Bayesian approach is close to Kriging in a manner
that a special construction is used to interpolate 
the values of the output of the deterministic code rather than the values of a random field 
and uncertainty intervals for untried values of inputs
are calculated; see Koehler, Owen (1996), Kennedy, O'Hagan (2001). 
One run by a computer model may require considerable time. Thus the main problem 
is to reduce an uncertainty of inferences on a computer model by making only a few runs.
Consequently, we are faced with the problem of optimal choice of experimental conditions.

The present paper is organized as follows. 
In Section 2 we review experimental designs for a multivariate case
in order to choose the most appropriate criteria of optimality. 
In Section 3 we propose a fast algorithm for constructing 
good optimal designs for computer experiments.

\section{Comparison of natural and computer experiments}

\vspace{-3.2mm}
Basic features of natural and computer experiments are presented in the following two-column 
style\footnotetext[1]{Without sparsity assumption we need a lot of runs to construct an unbiased 
 low uncertain predictor.}.

\vspace{0.1mm}
\noindent
\begin{tabular}{p{0.44\textwidth}c|cp{0.44\textwidth}}
 Natural experiments &\!\!&\!\!& Computer experiments \\
 \hline
 The response is observed with errors which may be correlated.  
 &\!\!&\!\!& The output is deterministic. The running of a computer code at the same inputs gives the same output.\\
 The response is described either by known regression function with unknown parameters
 or by multivariate linear or quadratic model  which is valid at a design subspace.
 &\!\!&\!\!& A computer code is considered to be like as a black box. 
 The main assumption is factor sparsity, that is
 the output depends in nonlinear way only on a few number of 
 inputs$^1$.\\
  A primary objective is to estimate parameters or find conditions 
  which maximize response.
  &\!\!&\!\!&  A primary objective is to fit a cheaper unbiased low uncertain predictor.  \\
  Other aims are identifying variables which have a significant effect, etc. 
  &\!\!&\!\!& Other aims are calibration of model parameters to physical data, optimization of output, etc.\\ 
  Optimal design is typically to minimize the (generalized) variance
  of estimated characteristics. 
  \newline
  Optimal designs are, for example, factorial, incomplete block, orthogonal, central composite,
  screening and $D$-optimal designs.
  &\!\!&\!\!& Optimality criteria is the minimization of mean square error
  over design space or the maximization of entropy. \newline 
  Optimal design is space-filling design.
  Latin hypercube design is recommended in many papers.\\ 
\end{tabular}
\vspace{0.1mm}

\noindent
Note that optimal designs for natural experiments mostly have two or three points 
in projection on each coordinate, 
e.g. the $2^{k-p}$ block and orthogonal designs have two points in projection,
central composite design has three points in projection. This fact is a consequence of
the multivariate linear or quadratic model which is assumed to be valid. 
Such designs are not suitable for computer experiments
since we assume that the output may be highly nonlinear in several variables.
Due to the objectives of computer experiments, optimal design  should minimize mean square error between
the prediction of response at untried inputs and the true output.
This criterion leads to the optimal design which should fill an entire design space uniformly
at the initial stage of computer experiments. The examples of space-filling design are
Latin hypercube design, sphere packing design, distance based design, uniform design,
design based on random or pseudo-random sequences, see Santner et al. (2003), Fang et al. (2006).
The optimal design should be a dense set in projection to each coordinate 
and should be a dense set in entire design space. 
Each of the above space-filling designs has attractive properties and satisfies
some useful criterion.  As far as is known, the best design should optimize a compound criterion.

\section{Latin Hypercube Designs}
 \vspace{-2mm}

At first, we need to recall an algorithm for construction of LH designs, 
which was introduced in McKay et al. (1979).
The algorithm generates $n$ points in dimension $d$ in the following manner.
1) Generate $n$ uniform equidistant points $x_{1}^{(s)},\ldots,x_{n}^{(s)}$ 
in the range of each input, $s=1,\ldots,d$.
2) Generate a matrix $(p_{i,j})$ of size $d\times n$ such that each row is 
a random permutation of numbers $1,\ldots,n$
and these permutations are independent.
3) Each column of the matrix $(p_{i,j})$ corresponds to a design point, that is
$(x_{p_{1,j}}^{(1)},\ldots,x_{p_{d,j}}^{(d)})^T$ is $j$th point of LHD.

Without loss of generality, we assume that the range of each input is $[0,1]$  and 
$x_{j}^{(s)}\in \mathcal{R}=\{0,1/(n-1),2/(n-1),\ldots,1\}$.

By construction, LHD has the best filling of range in projection on each coordinate.
Unfortunately, LHD may have a poor filling of entire hypercube.
Several criteria of optimality are introduced in order to choose a good LHD in a class of all LHD. 
Maximin criterion is a maximization of minimal distance
 \vspace{-2mm}
\bea
 \Psi_p(L)=\min_{i,j=1,\ldots,n} ||x_i-x_j||_p=
 \min_{i,j=1,\ldots,n} \left(\sum_{s=1}^d|x_{s,i}-x_{s,j}|^p\right)^{1/p}
\eea\vskip-2mm\noindent
usually used with $p=2$ where $x_i=(x_{1,i},\ldots,x_{d,i})^T$ is $i$th point of design $L$.
An LHD which maximize $\Psi_p(L)$ is called by maximin LHD.
Audze-Eglais criterion introduced in Audze, Eglais (1977) is a sum of 
forces between charged particles
and is a minimization of
 \vspace{-2mm}
\bea
 \Psi_{AE}(L)=\sum_{i=1}^n\sum_{j=i+1}^n \frac{1}{||x_i-x_j||^2_2}.
\eea\vskip-2mm\noindent
Others criteria of uniformity are star $L_2$-discrepancy, centered  $L_2$-discrepancy, 
wrap-around $L_2$-discrepancy which are motivated by quasi-Monte-Carlo methods
and the Koksma-Hlawka inequality, see Hickernell (1998), Fang et al. (2000).
Algorithms of optimization are studied in a number of papers, 
the local search algorithm in
Grosso et al. (2008),  the enhanced stochastic evolutionary algorithm in Jin et al. (2005), 
the simulated annealing algorithm in Morris, Mitchell, (1995)
the columnwise-pairwise procedure in Ye et al. (2000), 
the genetic algorithm  in Liefvendahl,  Stocki, (2006) and  Bates et al. (2003),
the collapsing method in  Fang, Qin (2003).
Cited authors concentrate on the case of low dimensions.

Basing on an analysis of papers on computer experiments, we can say that the size of LHD
is approximately equal to the input dimension multiplied by 10, that is $n\approx 10d.$
Further we propose a fast algorithm of constructing good LHD for the case
of high dimensions which is not studied, to the best of our knowledge.

First, we need to study features of random LHD generated by the above algorithm.
Let $L=\{x_1,\ldots,x_n\}$ be a LHD. Let $r_i$ be a minimal distance between $x_i$ and other points of $L$;
that is $r_i=\min_{j\neq i}||x_i-x_j||_2$
(further we consider euclidian distances). These distances characterize a design $L$.
Let $Q_\alpha$ denote an $\alpha$-percentile for sample $r_1,\ldots,r_n$.
Averaged values of low and upper quartiles, $Q_{0.25}$  and $Q_{0.75}$, are presented in table \ref{PW:luqrLHD}.
We see that the inter-point distances are varied and the quarter of distances are quite small.
Also note that distances between points is increased as the dimension is increased since $n=10d$.

\begin{table}[!hhh]
  \caption{\it Low and upper quartiles of distances between points of $n$-point random LHD 
  for different dimensions, $n=10d$.}
  \label{PW:luqrLHD}
  \smallskip
  \centering
  \begin{tabular}{|c| c|c|c|c| c|c|c|c| c|c|c|}
  \hline
  $d$       & 2   &3    &4    &5    &6    &7    \\
  \hline
  $Q_{0.25}$&0.108&0.167&0.232&0.305&0.368&0.434\\
  $Q_{0.75}$&0.175&0.270&0.347&0.431&0.502&0.573\\
  \hline
  \hline
  $d$       &8    &9    & 10  &14   &20   &\\
  \hline                                  
  $Q_{0.25}$&0.494&0.554&0.610&0.821&1.096&\\
  $Q_{0.75}$&0.636&0.699&0.757&0.972&1.249&\\
  \hline     
  \end{tabular}
\end{table}

For construction of $n$-point LHD with a given inter-point distance $r$
at dimension $d$, we propose the following heuristic algorithm.

\medskip
\goodbreak
{\bf Algorithm.}
\begin{enumerate}
 \vspace{-2mm}
  \item Let $L_k$ is a $k$-point design at $k$th step. Let $L_1=\{x_1\}$ where
  $x_1$ is a random point in the middle of $\mathcal{R}^d$ 
  such that its coordinates are unequal to each other.\vspace{-2mm}
  \item Compute a boolean matrix $B=\{B_{i,j}\}$ of size $d\times n$ upon $L_k$ such that $b_{i,j}=1$ ('used')
  if there exists a point in $L_k$ with $i$th coordinate which equals $(j-1)/(n-1)$, and 0 ('unused') otherwise.
 \vspace{-2mm}
  \item Generate a random point $z=(q_1,\ldots,q_d)^T/(n-1)\in\mathcal{R}^d$ such that each coordinate is unused;
  that is $b_{i,q_i}=0$, $i=1,\ldots,d$ except one random coordinate which should be taken nearby $0.5$.
 \vspace{-2mm}
  \item Create a set $C$ of candidate points in $\mathcal{R}^d$ with unused coordinates which are approximate points
  which are the closest and the furthest point from $z$ lied on spheres $S_r(x_j)$ 
  with centers $x_j\in L_k$ and radius $r$, $j=1,\ldots,k$.
 \vspace{-2mm}
  \item Find a point $x^*\in C$ such that $x^*$ lies outside of all $S_r(x_j)$, 
  that is $||x^{*}-x_j||>r$, $j=1,\ldots,k$. If there exist several such points, choose a point
  which minimizes $\#\{s:\sum_{i=1}^d B_{i,s}=m^*\}$, where $m^*=\min_{j=1,\ldots,n}\sum_{i=1}^d B_{i,j}$
  and $B=B(L_k\bigcup\{x^*\})$.
 \vspace{-2mm}
  \item Add $x^{*}$ to design, that is $L_{k+1}=L_k\bigcup\{x^*\}$. Stop at $n$th step.
 \vspace{-2mm}
  \item If we could not find $x^{*}$ at step 5, go to step 3. If we could find $x^*$ after several trials, 
  we should decrease $r$ since it is impossible to find a point which is far from $L_k$ at given distance $r$.
\end{enumerate}
 \vspace{-2mm}

\begin{table}[!hhh]
  \caption{\it Low and upper quartiles and $Q_{0.1}$ of distances 
  between points of $n$-point SLHD
  for different dimensions, $n=10d$. The value $r^*$ of maximin LHD is given.}
  \label{PW:lqoptLHD}
  \smallskip
  \centering
  \begin{tabular}{|c| c|c|c|c| c|c|c|c| c|c|c|}
  \hline
  $d$       & 2   &3    &4    &5    &6    &7    \\
  \hline
  $Q_{0.1}$ &0.217&0.310&0.351&0.393&0.512&0.584\\
  $Q_{0.25}$&0.217&0.312&0.363&0.476&0.535&0.617\\
  $Q_{0.75}$&0.217&0.323&0.409&0.486&0.552&0.626\\
  \hline
  $r^*$     &0.223&0.360&0.476&0.589&0.687&0.779\\
  \hline
  \hline
  $d$       &8    &9    & 10  &14   &20   &\\
  \hline                                   
  $Q_{0.1}$ &0.679&0.763&0.823&1.035&1.268&\\
  $Q_{0.25}$&0.694&0.765&0.824&1.037&1.271&\\
  $Q_{0.75}$&0.706&0.774&0.836&1.045&1.281&\\
  \hline                                   
  $r^*$     &0.867&0.950&1.021&-&-        &\\
  \hline     
  \end{tabular}
\end{table}

Let a design obtained by Algorithm be called  SLHD.
Numerical results show that Algorithm is fast and work well for any dimension.
It  requires 40 seconds to compute 100-point SLHD at dimension $d=10$
and 60 seconds for 140-point SLHD at $d=14$ 
and 200 seconds for 200-point SLHD at $d=20$ on PC 2.1GHz.
The choice of $r$ should be smaller  than $r^*$ 
where $r^*$ is the minimal distance between points of exact maximin LHD.
Since $r^*$ is unknown, we recommend the running Algorithm with different $r$, 
say, start with $Q_{0.75}$ for random LHD and increase it by small increment.
The decreasing of $r$ at step 7 does not mean that SLHD does not exist for given $r$
and is a consequence a poor placement of points at previous iterations.

\begin{figure}[!hhh]
\centering
\hspace{-10mm}\includegraphics[width=0.48\textwidth]{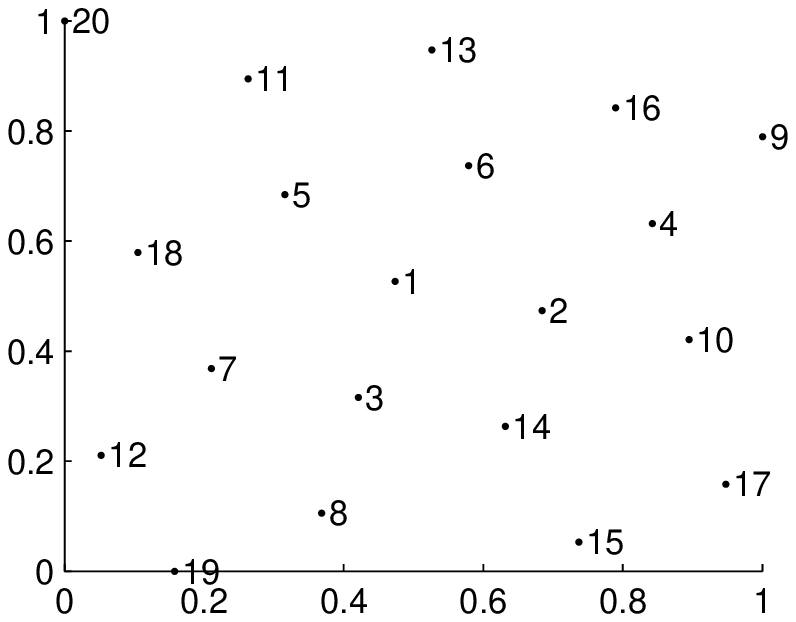}
\includegraphics[width=0.48\textwidth]{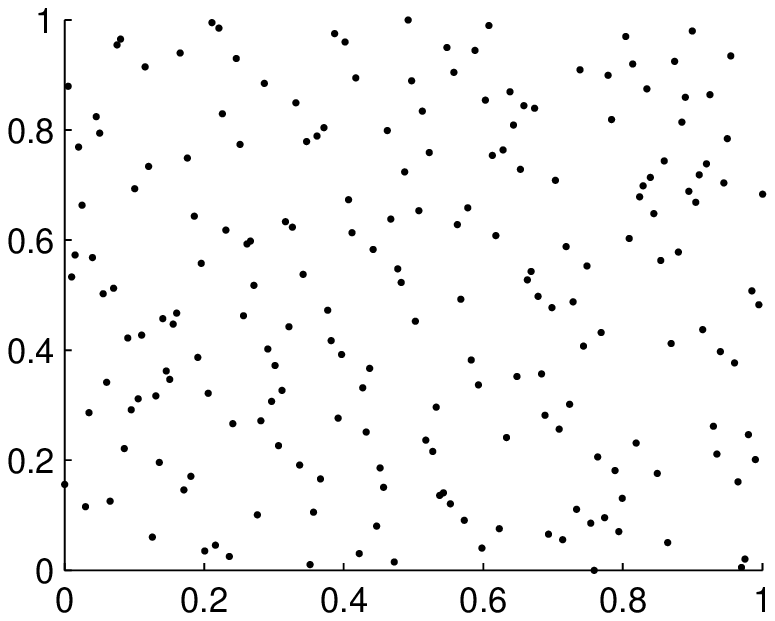}\kern-10mm
\caption{The points of SLHD for $d=2$ with the order of including (left) 
and two coordinates of points of SLHD for $d=20$ (right).}
\label{PW:fig:dim2-20}       
\end{figure}

Features of SLHD are  presented in Table \ref{PW:lqoptLHD}.
Values of $r^*$ are taken from web-site http://www.spacefillingdesigns.nl/.
We see that  90\% of inter-point distances of SLHD are higher than 
the most of distances at random LHD. 
Thus SLHD  has a better filling of entire design space.
Figure \ref{PW:fig:dim2-20} display points of SLHD  for $d=2$ and $d=20$.
We can see a quite uniform filling of square.
Further improvement of experimental design can be done by applying the local search 
or the simulated annealing algorithm.

\section{Conclusion}
The algorithm of construction of LHD with given inter-point distance is constructed and studied.
By the algorithm we can quickly compute LHD such that the most of inter-point distances are larger 
than distances at random LHD. The proposed algorithm is more efficient than simply generate 
many random LHDs and choose the best one.

\end{document}